\documentclass[twocolumn,aps,prl,superscriptaddress,showpacs,groupedaddress]{revtex4-1}
\usepackage{graphicx,dcolumn}
\usepackage{amsfonts,amssymb,amsmath,mathbbol,bbm}
\usepackage[english]{babel}
\usepackage{braket}
\newcommand{\nn}{\nonumber}

\begin{document}

\author{Tobias Geiger}
\author{Thomas Wellens}
\author{Andreas Buchleitner}
\affiliation{Physikalisches Institut, Albert-Ludwigs-Universit\"at
 Freiburg, D-79104 Freiburg, Germany}
 \title{Inelastic Multiple Scattering of Interacting Bosons in Weak Random Potentials}
\date{\today}

\begin{abstract}

We develop a diagrammatic scattering theory for
interacting bosons in a three-dimensional, weakly disordered potential. We show how collisional energy transfer between the bosons induces the thermalization of the inelastic single-particle current which, after only few collision events, dominates over the elastic contribution described by the Gross-Pitaevskii ansatz.

\end{abstract}


\maketitle

Bose-Einstein condensates, initially fascinating by themselves, have turned into a playground for a wide range of physical phenomena, reaching from condensed matter 
physics to cosmology \cite{lewenstein07,yasunari09}. A particularly interesting subject which recently spurred a lot of experimental and theoretical activities is the field of 
coherent many-particle quantum transport in disordered potential landscapes, due to exceptional experimental control on both, the 
confining potential as well as the inter-particle interactions \cite{modugno10}. This allows 
a detailed scrutiny of the hitherto largely unaccessible interplay of many-particle \cite{lewenstein07,mayer11,leboeuf10,haemmerling11} and 
disorder-induced \cite{modugno10} quantum transport phenomena, at an unprecedented level. In particular, given the precise knowledge of the microscopic constituents and interactions which
define the many-particle eigenstates and dynamics, these systems open new perspectives for an improved understanding of the emergence of collective and/or 
thermodynamic behavior from fundamental dynamical laws, with the thermalization problem as one central issue we will here embark on. Alternatively to various dynamical treatments of thermalization, in diverse physical contexts 
\cite{srednicki94, ponomarev06, rigol08, polkovnikov11,khatami12}, we formulate a linear, $N$-particle scattering scenario off a disordered potential. We will see that this lends itself to
 a transparent and physically intuitive understanding of the collision-induced thermalization of a quantum system, under rather general and experimentally easily accessible conditions. 
Furthermore, in the context of condensate dynamics in disordered potentials, our $N$-particle scattering approach defines a qualitative improvement over the wide-spread mean-field (or Gross-Pitaevskii) treatment, and a first step towards a linear scattering theory for interacting many particle systems.

Specifically, we consider an interacting bosonic gas scattering off a three dimensional, weakly disordered potential. We show that the thermalization is
mediated by {\em inelastic} contributions to the scattering amplitude, that amend and rapidly dominate over the strictly elastic, collective behavior described by the 
Gross-Pitaevskii equation. This establishes a unifying framework for ``condensate depletion" and the formation of a ``thermal cloud", 
as encountered e.g.~in \cite{ernst10},
under strictly unitary many-particle
evolution.  The r\^ole of the disorder is to randomize the individual particles' momenta, as necessary prerequisite for seeding inelastic collision events. 

While weak particle-particle interactions can be treated perturbatively under the assumption that the condensate be close to thermal equilibrium \cite{huang92,gaul11}, this is not
adequate any more in our present situation far from equilibrium. We therefore develop a diagrammatic theory which 
involves
a non-perturbative summation of 
all
those contributions which survive the 
average over the weakly disordered potential.

Let us start with a description of our scattering setup: 
Initially, each atom is prepared in the same single-particle momentum eigenstate with wave vector ${\bf k}_i$, pointing in $z$-direction. 
Then, the atoms enter a
three dimensional slab 
with thickness $L$ along the $z$-direction, and infinite extension in $x$- and $y$-direction. 
Within the slab,
the atoms experience scattering from a random potential $V$, and collisions 
due to particle-particle interaction $U$.
On exit from the slab, 
the
average spectral flux density $J_E$, i.e., the flux of particles with energy $E$ 
averaged over different realizations of the disorder potential $V({\bf r})$,
is detected.

Our microscopic scattering theory starts from an expansion of the $N$-particle scattering {\em amplitude} 
in powers of $V$ and  $U$. Each term in this expansion defines a scattering diagram and is composed of the following three elements: (i) the Green's function
$G_0$ for a single particle in free space, (ii) scattering of a single particle by the disorder potential $V({\bf r})$, and (iii) the two-particle $T$-matrix describing collisions
between particles:
\begin{equation}
\bra{{\bf k}_3,{\bf k}_4}\hat{T}\ket{{\bf k}_1, {\bf k}_2}=\delta_{{\bf k}_1+{\bf k}_2,{\bf k}_3+{\bf k}_4}t(k_{12})\,,\label{smatrix2}
\end{equation}
which, for a short-range interaction potential $U$ and the $s$-wave scattering approximation $a_s k_{1,2}\ll 1$, can be approximated by \cite{varenna99}:
\begin{equation}
t(k_{12})=16\pi\left(a_s-\frac{ik_{12}a_s^2}{2}+O(k_{12}^2a_s^3)\right)\,,
\label{tcoup}
\end{equation}
with $s$-wave scattering length $a_s$,  $k_1=|{\bf k}_1|,\,k_2=|{\bf k}_2|$ and $k_{12}=|{\bf k}_1-{\bf k}_2|$. The scattering length together with the density $\rho_0$ of particles in the initial state defines 
another length scale, $\ell_{\rm int}=1/8\pi a_s^2\rho_0$
-- the mean free path between two successive collision events. 
The probability that a third atom is located at distance $a_s$ from the colliding pair is assumed to be small, $a_s^3 \rho_0\ll 1$, 
such that three-particle collisions can be neglected \cite{fedichev96}. 

The $N$-particle scattering amplitude obtained by all combinations of the above building blocks (i -- iii)
defines the final state $|f\rangle=\hat{S}|i\rangle$, where $|i\rangle=|N{\bf k}_i\rangle$ is the initial state with all $N$ particles in mode ${\bf k}_i$. 
The measured 
flux $J_E=\langle f|\hat{J}_E|f\rangle$ 
is derived from the single-particle observable $\hat{J}_E$
which annihilates one particle, with 
the remaining $N-1$ particles traced over. To end up with a statistically robust quantity, one finally 
needs to average 
over 
different realizations of the disorder, 
which we assume Gaussian distributed, 
with mean 
$\overline{V({\bf r})}=0$,
and 
correlation 
\begin{equation}
\overline{V({\bf r}_1)V({\bf r}_2})=\frac{4\pi}{\ell_{\rm dis}}\delta({\bf r}_1-{\bf r}_2)\,,\label{2pc}
\end{equation}
where $\ell_{\rm dis}$ is the mean free path for scattering off the disorder potential. 
Under the further assumption of weak disorder, $k\ell_{\rm dis}\gg 1$,
only so-called {\em ladder diagrams} \cite{vanrossum99}, where the amplitude and the conjugate amplitude undergo the same sequence of scattering events,
survive the disorder average. 
$G_0$ is replaced by the 
average single-particle Green's function
\begin{equation}
G_E(k)=\frac{1}{\tilde{k}_E^2-k^2}\,,
\label{aspgf} 
\end{equation}
where $\tilde{k}_E=\sqrt{E}+i/2\ell_{\rm dis}$, in 
units of $\hbar^2/2m$.
\begin{figure}
\centerline{\includegraphics[width=8cm]{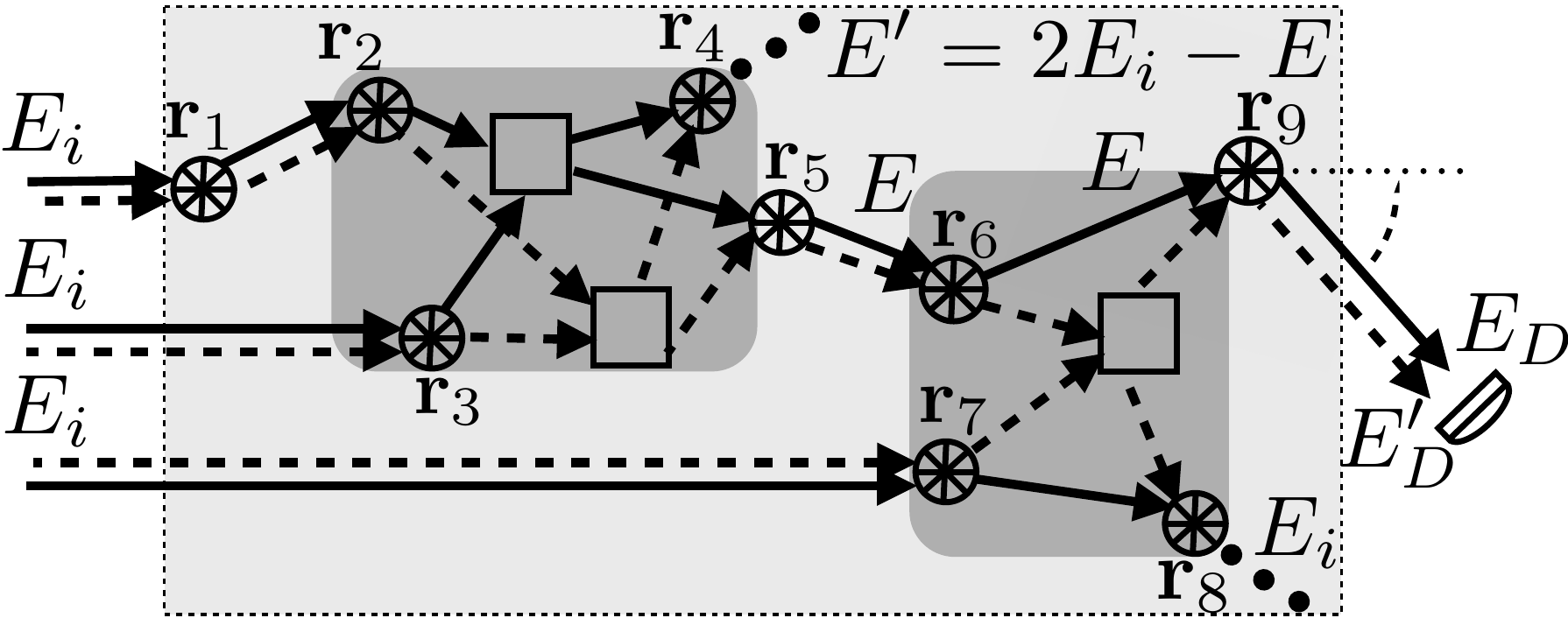}}
\caption{Example of a ladder diagram describing the propagation of three interacting particles 
in a slab with a random scattering potential. Pairs of conjugate amplitudes (solid and dashed arrows, respectively) undergo the same sequence of scattering events (encircled stars) 
induced by the disorder potential, at ${\bf r}_1,\dots, {\bf r}_9$. Due to particle-particle collision events (squares), the particles redistribute their energies. Here, 
solid and dashed arrows correspond to disorder averaged single-particle Green's functions (\ref{aspgf}) and their complex conjugates, respectively. Upon flux detection,
one particle is annihilated, 
while the undetected particles are traced over (dots).}
\label{ladder}
\end{figure}

A diagram contributing to the average flux is then constructed as follows: we take one diagram for the $N$-particle scattering amplitude, another one for the 
conjugate amplitude, group them together into a ladder diagram, detect one of the outgoing particles, and trace over the other ones. A typical example is shown in Fig.~\ref{ladder}.
Among the $N$-particle ladder diagrams thus constructed, we neglect all those where two particles which interacted 
once meet again. 
Alike the neglect of non-ladder diagrams, this 
approximation is valid for 
$k\ell_{\rm dis}\gg 1$, 
and allows us to trace over the undetected particles after their interaction with the detected particle, as 
shown in Fig.~\ref{ladder}.
Finally, we assume that at least one disorder scattering event occurs between two collision events. This is justified if $\ell_{\rm int}\gg\ell_{\rm dis}$, which, for a realistic 
scenario, is easily accessible by adjusting $\ell_{\rm dis}$, i.e., the disorder strength, accordingly.

\begin{figure}
\centerline{\includegraphics[width=8.6cm]{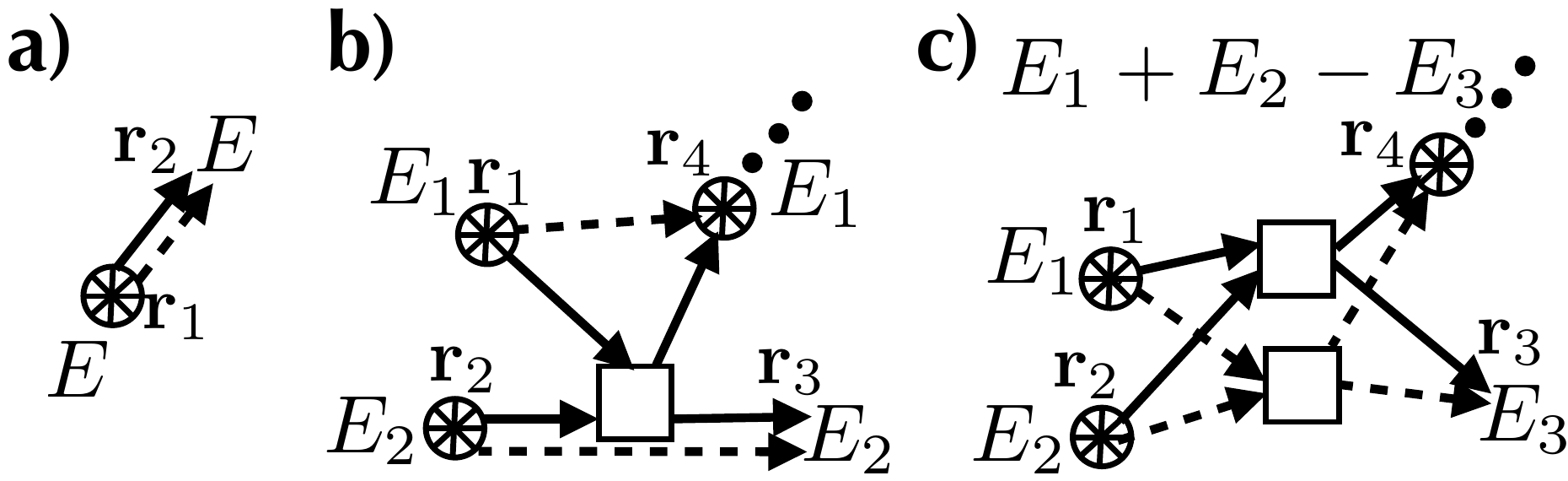}}
\caption{The three building blocks from which all ladder diagrams (see Fig.~\ref{ladder}) are constructed. (a) Single-particle propagation in the disorder potential, see Eq.~(\ref{avint}). (b) Nonlinear elastic scattering $g_{E_1;E_2}$. (c) Nonlinear inelastic scattering $f_{E_1,E_2;E_3}$, see Eq.~(\ref{f}).}
\label{nonlin}
\end{figure}

Under these assumptions, any diagram contributing to the particle flux is composed 
of three building blocks, see Fig.~\ref{nonlin}.
The first one, Fig.~\ref{nonlin}(a), denotes scattering of a single particle off the disorder potential at ${\bf r}_1$, and subsequent propagation to the next scattering event at ${\bf r}_2$:
\begin{equation}
P_E({\bf r}_1,{\bf r}_2)=\frac{4\pi}{\ell_{\rm dis}}\left|\sum_{\bf k}e^{i {\bf k}{\bf r}_{12}} G_E(k)\right|^2=\frac{e^{-r_{12}/\ell_{\rm dis}}}{4\pi\ell_{\rm dis}r_{12}^2}\label{avint}\,,
\end{equation}
where ${\bf r}_{12}={\bf r}_1-{\bf r}_2$.
In the second building block, Fig.~\ref{nonlin}(b), one pair of amplitudes (solid lines) exhibits a particle-particle collision event, whereas the 
other pair (dashed lines) does not experience a collision. Consequently, the energies $E_1$ and $E_2$ of both particles remain conserved (otherwise the solid and dashed amplitudes could not be grouped together). The diagram Fig.~\ref{nonlin}(b) hence 
represents a {\it nonlinear elastic} scattering contribution, which we denote by 
$g_{E_1;E_2}({\bf r}_1,{\bf r}_2,{\bf r}_3)$. 
Furthermore, one can show that, if one neglects the second order term $k_{12}a_s^2$ in the two-particle scattering amplitude, Eq.~(\ref{tcoup}), this diagram is equivalent to a diagram obtained from the stationary  Gross-Pitaevski equation \cite{wellens09a}. In contrast, the diagram Fig.~\ref{nonlin}(c) 
represents {\it inelastic} scattering events, {\it not} accounted for by the Gross-Pitaevski equation:  It describes a collision between two particles, where the energies of both particles change from $E_1$ and $E_2$ to $E_3$ and $E_4=E_1+E_2-E_3$, respectively. The 
weight of such processes is given by:
\begin{align}
&f_{E_1,E_2;E_3}({\bf r}_1, {\bf r}_2;{\bf r}_3)=2\times (16\pi a_s)^2\times\left(\frac{4\pi}{\ell_{\rm dis}}\right)^3\nn\\
&\times\sum_{{\bf k}}\delta(k^2-E_4)\int {\rm d}{\bf r}_4\Biggl| \frac{1}{2}\sum_{{\bf k}_1,{\bf k}_2,{\bf k}_3} e^{i({\bf k}_{1}{\bf r}_{41}+{\bf k}_{2}{\bf r}_{42}-{\bf k}_{3}{\bf r}_{43})}\nn\\
& \times G_{E_1}(k_1)G_{E_2}(k_2)G_{E_3}(k_3)G_{E_4}(|{\bf k}_1+{\bf k}_2-{\bf k}_3|)\Biggr|^2 \, . \label{f}
\end{align}
Note that 
Eq.~(\ref{f}) is 
quadratic 
in the small parameter $a_s$, i.e., inserting 
the first order 
contribution to $t(k_{12})$ in 
Eq.~(\ref{tcoup}) suffices.  The first factor 2 in Eq.~(\ref{f}) originates from the fact that  the solid and dashed  incoming amplitudes can be grouped  together in two different ways. It can be shown that this accounts for fluctuations of the atomic density inside the disordered slab \cite{wellens09b}. The sum over ${\bf k}$ represents the trace over the undetected particle. The factor $1/2$ in front of the second sum indicates that this sum is taken over the  subspace of symmetrized two-particle states $|{\bf k}_1,{\bf k}_2\rangle$.

We can now write down 
a nonlinear
integral equation for the average particle density at energy $E$:
\begin{align}
&I_E({\bf r}) = I_0({\bf r})\delta(E-E_i)+\int {\rm d}{\bf r}^\prime P_E({\bf r}^\prime,{\bf r})I_E({\bf r}^\prime)+\label{preint}\\
&+\int {\rm d}E_1~\left[g_{E_1;E} I_E({\bf r})+\int{\rm d}E_2~f_{E_1,E_2;E}I_{E_2}({\bf r})\right]I_{E_1}({\bf r})\,,\nn
\end{align}
which upon iteration produces all possible combinations of the above three building blocks,
non-perturbatively in the collision contributions $g_{{E_1};E}$ and $f _{E_1,E_2;E}$.
Here,
$I_0({\bf r})=\rho_0\exp(-z/\ell_{\rm dis})$ denotes the density $\rho_0$ of particles in the initial mode, attenuated by the propagation to position $z$ inside the slab.
Furthermore, we have employed a 
contact approximation for the collision terms, i.e.,  $g_{E_1;E}({\bf r}_1,{\bf r}_2,{\bf r}) \simeq \delta({\bf r}_1-{\bf r})\delta({\bf r}_2-{\bf r})g_{E_1;E}$, with
$g_{E_1;E}=\int{\rm d}{\bf r}_1{\rm d}{\bf r}_2~g_{E_1;E}({\bf r}_1,{\bf r}_2,{\bf r})$, and similarly for $f$. This is justified since we assume 
$\ell_{\rm int}\gg\ell_{\rm dis}$, and hence the spatial transport of particles between two 
points ${\bf r}^\prime$ and ${\bf r}$ is dominated by the single-particle propagator $P_E({\bf r}^\prime,{\bf r})$. With this approximation, 
Eq.~(\ref{f}) implies, with $k\ell_{\rm dis}\gg 1$:
\begin{equation}
\label{fex}
f _{E_1,E_2;E} =\frac{8\pi \ell_{\rm dis}a_s^2}{\sqrt{E_1 E_2E}} \left\{\begin{array}{cl}\sqrt{E} & E<E_1\\ \sqrt{E_1} & E_1\leq E\leq E_2\\ \sqrt{E_1+E_2-E} & E>E_2\end{array}\right.\,,
\end{equation}
for $E_2>E_1$. The expression for $g_{{E_1};E}$ can be calculated in a similar way from 
diagram Fig.~\ref{nonlin}(b). Alternatively, it can be extracted from Eq.~(\ref{fex}) and one of the two conditions
$\sqrt{E_2}g_{E_1;E_2}=-\int_0^{E_1+E_2} {\rm d}E~\sqrt{E} f_{E_1,E_2;E}$ or
$(E_1+E_2)\sqrt{E_2}g_{E_1;E_2}=-\int_0^{E_1+E_2} {\rm d}E~2E\sqrt{E} f_{E_1,E_2;E}$,
which guarantee conservation of the particle and energy flux, respectively, in Eq.~(\ref{preint}). In other 
words, inelastic scattering goes along with a corresponding reduction of the nonlinear elastic component. To this 
end, it is crucial to keep the second order term in Eq.~(\ref{tcoup}) in the expression for $g$, since the first-order term (i.e. the result 
predicted by the Gross-Pitaevski equation) vanishes within the ladder approximation \cite{wellens09a}.

\begin{figure}
\centerline{\includegraphics[width=8.4cm]{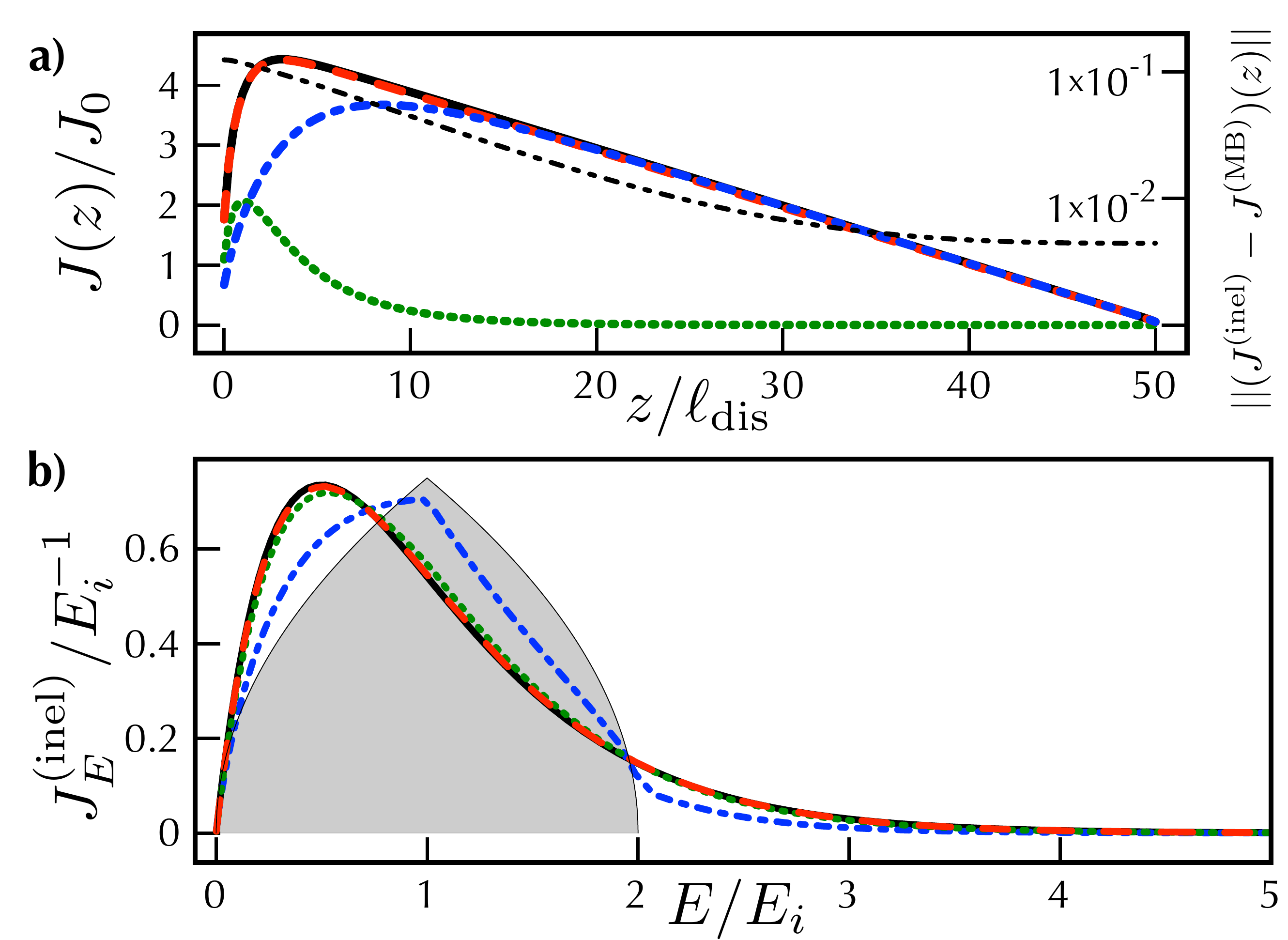}}
\caption{(Color online) (a) Flux density $J$ vs. position $z$ in a slab of thickness $b=L/\ell_{\rm dis}=50$, for
$\ell_{\rm dis}/\ell_{\rm int}=1/250$. Elastic (green, dotted) and inelastic (blue, dash-dotted) component add up to the total, interacting many particle flux density (black, solid), which coincides with the linear flux (red, dashed). The black dash-double-dotted line indicates the deviation of $J(z)$ from a Maxwell-Boltzmann flux density (y-scale on the right).
(b) Normalized energy distribution $J_E^{({\rm inel})}(z)$ of inelastically scattered atoms at $z=0$ (blue, dash-dotted), $z=L/3$ 
(green, dotted) and $z=L$ (red, dashed) within the slab; otherwise the same parameters as in (a). The distribution $f_{E_{\rm i},E_{\rm i};E}$ 
after one single inelastic scattering event (gray-shaded area), 
Eq.~(\ref{fex}), 
evolves into a Maxwell-Boltzmann distribution (black, solid) with average energy $E_{\rm i}$, upon penetration into the slab.}
\label{sfd}
\end{figure}

With these premises, we can now infer the total average flux $J(z)=\int {\rm d}E J_E(z)$ at position $z\in [0;L]$ within the slab, with the 
energy-dependent flux $J_E(z)=\sqrt{E}I_E(z)$,
in units of the incident flux $J_0=\sqrt{E_i}\rho_0$.
Here, $I_E(z)$ is obtained numerically via iterative solution of Eq.~(\ref{preint}).
Fig.~\ref{sfd}(a) shows the result 
for a slab 
thickness $b=L/\ell_{\rm dis}=50$, and weak interaction $\ell_{\rm dis}/\ell_{\rm int}=1/250$.
$J(z)$ exhibits the characteristic linear decay of diffusive (or Ohmic) transport \cite{akkermans07}, and 
equals the {\em linear} flux which is 
obtained from Eq.~(\ref{preint}) when setting $a_s=0$.
This
is due to the 
condition $\ell_{\rm int}\gg \ell_{\rm dis}$, and the corresponding 
contact approximation mentioned above,  together with 
the fact that, 
for the 3D white-noise potential (\ref{2pc}),
$\ell_{\rm dis}$ is independent of the particle's energy \cite{akkermans07}. In 
contrast to the linear case, however, 
$J(z)$ splits into an elastic and an inelastic component, defined by $J_E(z)=
J_E^{\rm (el)}(z)\delta(E-E_i)+J_E^{\rm (inel)}(z)$. We see that, in spite of the weakness of the interaction, the inelastic component rapidly dominates as the particles penetrate 
into 
the slab. This can be explained by the large number $b^2$ of (disorder) scattering events required to traverse a slab with thickness $b$. The 
expected number of two-body collision events correspondingly scales 
as $b^2 \ell_{\rm dis}/\ell_{\rm int}$.
By the same argument, three-body collisions can be neglected if $a_s^3\rho_0\ll \ell_{\rm int}/(\ell_{\rm dis}b^2)$. 
Note that the continuous emergence of an 
inelastic component of the flux, as described by our present, microscopic and strictly unitary treatment, is tantamount to the formation of what is colloquially called 
a ``non-condensed fraction" or ``thermal cloud",
since an $N$-fold product of a single-particle state (as required from the formal definition of a condensate via the stationary one-particle density matrix \cite{lieb05}) 
with fixed total energy implies fixed energies also for the individual particles. 

The normalized energy distribution $J_E^{(\rm inel)}(z)$ of the inelastic component is shown in Fig.~\ref{sfd}(b).
We see that, starting out from a distribution which is centered around $E_{\rm i}$ after the first inelastic event, the energy distribution approaches a Maxwell-Boltzmann distribution $J_E^{(\rm MB)}=4 E\exp(-2E/E_{{\rm i}})/E_{{\rm i}}^2$, with the 
average energy (or ``temperature") fixed by the incident particle energy $E_{\rm i}$. Note that due to the strongly reduced density, collisions become very unlikely towards the end of the slab, and hence the spectral inelastic flux densities between slab positions $z=L/3$ and $z=L$ are only slightly altered, Fig.~\ref{sfd}(b). This also manifests in Fig.~\ref{sfd}(a), by the saturation of the difference $||(J^{(\rm inel)}-J^{(\rm MB)})(z)||=\sqrt{E_{\rm i}\int d{\rm E}\left(J_E^{(\rm inel)}-J_E^{(\rm MB)}\right)^2(z)}$ between the inelastic and Maxwell-Boltzmann flux density, as a function of position in the slab. 

In summary, we 
formulated a microscopic 
transport theory for interacting bosons propagating in a random potential. We showed that the disorder-averaged single-particle density 
matrix relaxes to a stationary state which, after only few ($b^2 \ell_{\rm dis}/\ell_{\rm int}\approx 10$) collision events
inside the scattering region, coincides with a thermal Maxwell-Boltzmann distribution with ``temperature" given by the incident particles' energy.

For interacting particles with confinement rather than disorder, an analogous result was derived from a random matrix argument, under the constraint $1/k\leq a_s\ll\rho_0^{-1/3}$ \cite{srednicki94}.
Observe that the assumptions required for our theory, $k\ell_{\rm dis}\gg 1$ (weak disorder), $\ell_{\rm int}=1/(8\pi a_s^2\rho_0)\gg \ell_{\rm dis}$ (collisions less frequent than disorder scattering), and the $s$-wave scattering 
approximation, can be summarized as $a_s\ll 1/k\ll \ell_{\rm dis}\ll \ell_{\rm int}$. 
Since $k$ and $\ell_{\rm dis}$ may be adjusted by choosing the initial energy and the disorder potential appropriately,
our results hold for typical experimental parameters of ultracold bosonic gases, $\rho_0\approx (10^{18}-10^{21})\rm m^{-3}$ and $a_s\approx (10^{-8}-10^{-9})\rm m$  \cite{pethick08}, corresponding to
$\ell_{\rm int}\approx (4\times 10^{-2} - 4\times 10^{-7})\rm m$. 
We verified,
for a wide range of optical thicknesses $b$ and ratios $\ell_{\rm dis}/\ell_{\rm int}$, that Eq.~(\ref{preint}) indeed provides unique stationary solutions
and predicts full quantum thermalization of the flux density if the number of inelastic collisions exceeds $b^2 \ell_{\rm dis}/\ell_{\rm int}\approx 10$.
In fact, the collision terms in the transport equation (\ref{preint}) exactly reproduce Boltzmann's kinetic equation for a gas of 
classical particles, for which the stationary energy distribution is known to be given by $J_E^{(\rm MB)}$ \cite{huang87}. 
We note that rigorous derivations of a 
nonlinear quantum Boltzmann equation similar to the collisional terms in  Eq.~(\ref{preint}) have been attempted recently \cite{spohn07, benedetto08}, though
in the absence of a random 
potential. It is precisely the presence of the latter, however, which allows for a rigorous quantification of the regime of validity of
Eq.~(\ref{preint}), in terms of the parameters $k$, $\ell_{\rm dis}$ and $\ell_{\rm int}$, in our present treatment.

Let us conclude with a discussion of possible extensions of 
our theory:
Since the main idea -- neglect of all but ladder diagrams for weak disorder -- is not restricted to stationary scattering processes, 
we expect that the present theory can be extended to time-dependent scenarios such as, e.g., expansion of an initially confined 
condensate in a random potential \cite{cherroret11}. Furthermore, relaxing the contact approximation (assuming $\ell_{\rm int}\gg \ell_{\rm dis}$) for the 
collision terms allows to enter a regime of stronger interactions where, e.g., repulsion or attraction between particles will affect the spatial density profile. 
Finally, by means of crossed diagrams, one can study interference phenomena like coherent backscattering \cite{hartung08}, and clarify how these are modified by interactions.

We thank Pierre Lugan for fruitful discussions and a critical reading of the manuscript,
and acknowledge partial support by DFG research unit FG760. T. G. acknowledges funding through DFG grant BU1337/8-1.

\end{document}